\documentclass[english,aps,manuscript]{revtex4}
\usepackage[T1]{fontenc}
\usepackage[latin1]{inputenc}
\usepackage{amsmath}
\usepackage{amssymb}

\makeatletter
\usepackage{amssymb}

\usepackage{graphicx}
\usepackage{dcolumn}
\usepackage{bm}
\usepackage{amsfonts}%
\usepackage{babel}

\usepackage{babel}
\makeatother
\begin{document}

\title{Exact Entanglement Cost of Multi-Qubit Bound Entangled States}

\author{Somshubhro Bandyopadhyay}

\email{som@qis.ucalgary.ca}

\affiliation{Institute for Quantum Information Science, University of Calgary,
Calgary, AB T2N 1N4, Canada. }

\author{Vwani Roychowdhury}

\email{vwani@ee.ucla.edu}

\affiliation{Electrical Engineering Department, UCLA, Los Angeles, CA 90095}

\begin{abstract}
We report the \emph{exact} entanglement cost of a class of multiqubit
bound entangled states, computed in the context of a universal model
for multipartite state preparation. The exact amount of entanglement
needed to prepare such states are determined by first obtaining lower
bounds using a cut-set approach, and then providing explicit local
protocols achieving the lower bound. 
\end{abstract}
\maketitle

\section{\textbf{Introduction}}

Quantum entanglement \cite{entanglement} has emerged as a new type
of physical resource, and is the key ingredient for quantum information
processing (QIP) tasks, including teleportation \cite{teleportation},
super dense coding \cite{BW92}, and secure key distribution \cite{E91}.
A physical resource (e.g., heat energy) is typically characterized
by a measure or unit, so that a given system could be assigned a physically
meaningful estimate of how much of the resource it contains (e.g.,
BTUs or calories). The purpose of defining such measures is operational
and practical: for example, if we are given an amount of fuel with
say 10,000 BTUs of recoverable heat energy, then one can derive a
quantitative estimate of how large a space it can heat and what should
be the efficiency of the conversion process. One could ask analogous
questions for entanglement: is there a measure of entanglement such
that one can use it to convey how many units of entanglement one can
extract from a given state, or how many units would one require to
prepare it. The critical importance of defining such relevant units
of entanglement was recognized early on by the QIP community and a
considerable amount of effort has been put into it \cite{BDSW95,VW01,VPRK97}.
Such efforts have met with only mixed success, and the case of multipartite
entanglement has proven to be particularly difficult. 

In the standard model for entanglement transformation, a state is
shared by a set of spatially separated parties, and is manipulated
via local operations and classical communications (LOCC). For the
case of \emph{bipartite} states, one can indeed define a physically
informative unit referred to as the \emph{ebit}, where one ebit stands
for the entanglement content of a singlet state. Given a bipartite
state, the \emph{entanglement cost} of a state is the number of singlets
required to prepare the state asymptotically, and \emph{entanglement
of distillation} is the amount of pure entanglement that can be extracted
per copy, also in the asymptotic sense \cite{BDSW95,Rains,hayden00}
can be suitably expressed in terms of ebits. 

Multipartite quantum systems, on the other hand pose considerable
difficulty in defining and computing a reasonable measure for entanglement.
For instance, given a quantum state comprising of $N$ subsystems,
one can group the subsystems into $2\leq M<N,$ groups, where some
of the subsystems can be considered to be a joint subsystem in a larger
Hilbert space\emph{. Entanglement} of each such $M$-partition is
in general different, and indeed, a state which may be separable under
certain partitions might be entangled in other partitions. However,
for GHZ class of states, the reduced entropy as in pure bipartite
states has been shown to be an appropriate entanglement measure \cite{bennett99}.
Also a geometric measure of entanglement, originally proposed for
bipartite systems, has been generalized for the case of multipartite
entangled states \cite{shimoni95,barnum01,wei03}. The measure tries
to estimate the distance in the Hilbert space from the closest separable
states, and provides a lower bound for entanglement of formation \cite{wootters98}.
This bound has been computed \cite{weigoldbart03} for the unlockable
bound entangled states of Smolin \cite{smolin00} and Dur \cite{dur00}.
As for the \emph{entanglement cost of multipartite states}, there
is no proposed general measure that can be reasonably defined and
accurately estimated for a large enough class of states. $\,\,$

This letter provides \emph{an exact entanglement cost of preparing
a class of multipartite bound-entangled states} recently introduced
in Ref \cite{BCRS04,AH04}. The entanglement costs of these states
are computed in the context of a universal model for multipartite
state preparation: the minimum total bipartite entanglement required
to prepare the given states via local operations and classical communications
(LOCC) and starting from an initial state comprising only pairwise
shared bipartite entanglements. The exact costs are determined by
first computing lower bounds using a cut-set approach, and then providing
explicit protocols for preparing the multipartite states (i.e., via
LOCC and pairwise shared bipartite entanglements among the parties)
that use the same total entanglement as the lower bounds. 

There are several implications of the results in this letter worth
noting: (i) To our knowledge, the exact entanglement cost of any mixed
multipartite entangled states, be it distillable or bound entangled,
has not been reported so far in the literature. These are first known
exact entanglement cost of multipartite mixed states. (ii) For bipartite
systems, the question whether the asymptotic entanglement cost per
copy can become zero for a bound entangled state has been resolved
recently \cite{vidalcirac01}. The question however was open for multipartite
states, and we answer it by showing that the entanglement cost of
a multipartite bound-entangled state does not approach zero in general
in the asymptotic limit. (iii) It has been demonstrated that in bipartite
systems, asymptotic manipulation is more efficient than single copy.
Our result shows that in multipartite case, even if the state involved
is a mixed one, asymptotic manipulation may not be more efficient
than single copy.

\section{\textbf{A Universal Model For Computing the Entanglement Cost of
Multipartite States}}

For bipartite systems, any state preparation involves pre-shared entanglement
between the two parties which is then manipulated via LOCC to prepare
the state in question. This can be suitably generalized for multipartite
systems as well. A universal model for the preparation of multipartite
states can be described as follows: The spatially separated parties
start with pair-wise and independent bipartite entanglement. The parties
then use LOCC to prepare the desired state. This universal model provides
a unique means of computing entanglement cost for multipartite states:
the sum of all the pairwise ebits used in the \textit{optimal} preparation
of the state. More precisely, for multi-qubit systems, let $\rho$
be a $N$-qubit state to be prepared. Suppose now $\rho_{B}$ be another
$N$ qubit state comprising only of pairwise bipartite entanglements
having the form: $\rho_{B}=\otimes\prod_{i<j}^{N}\sigma_{ij}$, where,
the index $ij$ refers to the pair of parties $(i,j)$ sharing the
state $\sigma_{ij}.$ Let the distillable entanglement between every
pair of parties, $(i,j)$ be $e_{ij}$ measured in ebits. It is clear
that there always exists a $\rho_{B}$ such that $\rho$ can be prepared
from $\rho_{B}$ via LOCC \emph{(in general under asymptotic manipulations}),
i.e., $\rho_{B}{}^{\otimes n}\rightarrow\rho^{\otimes m}$ where $\frac{m}{n}\rightarrow constant$
as $n\rightarrow\infty$. The entanglement cost, $E_{C}(\rho)$, is
given as: \begin{equation}
E_{C}(\rho)=\min\left(\sum_{i<j}e_{ij}\right)\,\,.\label{def:eq}\end{equation}
 That is, the \emph{entanglement cost} of a multipartite state is
the sum of the bipartite entanglement between the parties \emph{minimized
over all possible strategies for preparing the state}. The above definition
has a straightforward generalization for a general $N$ party system
where the $i-th$ party holds a quantum system of dimension $d_{i}$.
$$. $$

\begin{center}\textbf{A cut-set (C-S) approach to Estimate lower bounds
on $E_{C}$}\end{center}

$$ While
solving the above minimization problem is not an easy task in general,
calculating non-trivial lower bounds is a lot more straightforward.
We can use the well-known truism that \emph{distillable entanglement
across a bipartite cut cannot increase under LOCC.} If for example,
in a two-party configuration of a multipartite state one can distill
$n$ ebits of entanglement between the two groups, then while preparing
the entangled state in question, when all the relevant parties are
spatially separated, one must have had spent at least $n$ ebits of
entanglement across the same cut. Thus, the sum of the pairwise bipartite
entanglements crossing the cut must be at least $n$ ebits. By computing
distillable entanglement across all possible (or even a subset) of
the bipartite cuts, one can always obtain a lower bound on the amount
of entanglement that one needs to spend in preparing the state. 

\noindent \textbf{Remark 1:} Note that since distillable entanglement
can only increase in the asymptotic case, the lower bound on the cost
is \emph{independent of whether the distillable entanglements}, used
in computing the bound, \emph{correspond to the few-copy or the asymptotic
case}. Hence, any lower bound using the cut-set approach is also a
\emph{lower bound} in an \emph{asymptotic} sense. 

\begin{center}\textbf{Achieving the lower bound }\end{center}

The above two observations lead to the \textit{\emph{following strategy
adopted in this}} work: \textbf{(i)} Compute a lower bound on the
entanglement cost of a given multipartite state, based on the distillable
entanglement across different possible partitions, and \textbf{(ii)}
then, search for a strategy to prepare the given state using LOCC
on another state which consists of only pairwise shared bipartite
entanglement (i.e., ebits) such that the total bipartite entanglement
equals the lower bound computed in step (i). 

\noindent \textbf{Remark 2:} If we find a state consisting of pairwise
shared ebits from which (\emph{even if it is the single-copy case})
we can prepare the given multipartite state using LOCC, and the total
bipartite entanglement equals the \emph{lower bound} computed in step
(i), then the bound from step (i) is also the \emph{exact entanglement
cost} of the given multipartite state. That is, \emph{even though
the state is prepared using a single copy} of a state comprising pairwise
bipartite entanglement, one \emph{cannot do any better using asymptotic
manipulations}. This is because, the lower bound is already in the
asymptotic sense (see Remark 1), and hence, if one could use less
overall bipartite entanglement using asymptotic manipulations, then
it will lead to contradiction. 

We now use the above strategy to calculate the \emph{exact entanglement
cost} of a general class of \emph{multipartite bound-entangled states}.

\section{\textbf{Entanglement cost for Multi-Qubit Bound Entangled States}}

Recall that a multipartite quantum state is said to be bound entangled
if there is no distillable entanglement between any subset as long
as \textit{all} the parties remain spatially separated from each other.
If for such a state, entanglement can be distilled between two parties
by bringing a subset of the other parties together, then the state
is said to be an \emph{activable bound entangled} (ABE) state. We
now briefly describe a class of Bell-correlated ABE (BCABE) states,
introduced in \cite{BCRS04,AH04}. First, however, we introduce few
notations. The customary two qubit Bell states are defined as follows:
\begin{equation}
\left|\Phi^{\pm}\right\rangle =\frac{1}{\sqrt{2}}\left(\left|00\right\rangle \pm\left|11\right\rangle \right),\left|\Psi^{\pm}\right\rangle =\frac{1}{\sqrt{2}}\left(\left|01\right\rangle \pm\left|10\right\rangle \right)\,\,.\end{equation}

Consider now a system comprising of $2N,N\geq2$ qubits. Let $\left|p_{i}\right\rangle =\left|a_{1}^{i}a_{2}^{i}...a_{2N}^{i}\right\rangle $
where $a_{1}^{i}=0$, and $a_{j}^{i}\in\{0,1\}$, for all $j=2,\cdots,2N$
such that there is an even number of 0s in the string $a_{1}^{i}a_{2}^{i}...a_{2N}^{i}.$
Likewise, let $\left|q_{i}\right\rangle =\left|b_{1}^{i}b_{2}^{i}...b_{2N}^{i}\right\rangle ,$
where $b_{1}^{i}=0,$ and $b_{2}^{i},...,b_{2N}^{i}$ are either 0
or 1 with odd number of 0s in the string $b_{1}^{i}b_{2}^{i}...b_{2N}^{i}$.
One can also define the states orthogonal to $\left|p_{i}\right\rangle ,\left|q_{i}\right\rangle $
as: $\left|\overline{p_{i}}\right\rangle =\left|\overline{a_{1}^{i}}\overline{a_{2}^{i}}...\overline{a_{2N}^{i}}\right\rangle $
and $\left|\overline{q_{i}}\right\rangle =\left|\overline{b_{1}^{i}}\overline{b_{2}^{i}}...\overline{b_{2N}^{i}}\right\rangle $
where $\left\langle \overline{a_{j}^{i}}|a_{j}^{i}\right\rangle =0=$
$\left\langle \overline{b_{j}^{i}}|b_{j}^{i}\right\rangle ,\forall j=1,...,2N$
and $i=1,...,2^{2N-2}.$ Note that the four sets of states, defined
by $\left|p_{i}\right\rangle $'s, $\left|\overline{p_{i}}\right\rangle $'s,
$\left|q_{i}\right\rangle $, and $\left|\overline{q_{i}}\right\rangle $'s
respectively, are non-overlapping and all have same cardinality, and
they together span the complete Hilbert space of $2N$ qubit systems. 

Now define the cat or GHZ basis:\begin{equation}
\left|\Phi_{i}^{\pm}\right\rangle =\frac{1}{\sqrt{2}}\left(\left|p_{i}\right\rangle \pm\left|\overline{p_{i}}\right\rangle \right),i=1,...,2^{2N-2}\label{eq:phiplusminus}\end{equation}
 \begin{equation}
\left|\Psi_{i}^{\pm}\right\rangle =\frac{1}{\sqrt{2}}\left(\left|q_{i}\right\rangle \pm\left|\overline{q_{i}}\right\rangle \right),i=1,...,2^{2N-2}\label{eq:psiplusminus}\end{equation}
 We will use the notation $\left[\cdot\right]$ for pure state projector
$\left|\cdot\right\rangle \left\langle \cdot\right|$. Let us now
define the following \textit{2N} qubit density matrices:\begin{equation}
\rho_{2N}^{\pm}=\frac{1}{2^{2N-2}}\sum_{i=1}^{2^{2N-2}}\left[\Phi_{i}^{\pm}\right],\sigma_{2N}^{\pm}=\frac{1}{2^{2N-2}}\sum_{i=1}^{2^{2N-2}}\left[\Psi_{i}^{\pm}\right]\end{equation}
 In \cite{BCRS04} an interesting recursive relation was derived relating
bound entangled states of \textit{2N-2} qubits with that of \textit{2N}
qubits: 

\begin{equation}
\rho_{2N}^{\pm}=\frac{1}{4}(\left[\Phi^{+}\right]\otimes\rho_{2N-2}^{\pm}+\left[\Phi^{-}\right]\otimes\rho_{2N-2}^{\mp}+\left[\Psi^{+}\right]\otimes\sigma_{2N-2}^{\pm}+\left[\Psi^{-}\right]\otimes\sigma_{2N-2}^{\mp})\label{rhoplusminus}\end{equation}

\begin{equation}
\sigma_{2N}^{\pm}=\frac{1}{4}(\left[\Psi^{+}\right]\otimes\rho_{2N-2}^{\pm}+\left[\Psi^{-}\right]\otimes\rho_{2N-2}^{\mp}+\left[\Phi^{+}\right]\otimes\sigma_{2N-2}^{\pm}+\left[\Phi^{-}\right]\otimes\sigma_{2N-2}^{\mp})\label{sigmaplusminus}\end{equation}

The class of states $\rho_{2N}^{\pm},\sigma_{2N}^{\pm}$ have been
shown to be activable bound entangled in Ref \cite{BCRS04}. Let us
just note that the states are bound entangled when all $2N$ parties
are separated from each other. This is the configuration where the
entanglement cost will be evaluated. Furthermore the set of states
are connected to each other by local pauli operations on one qubit.
Here we also note that if any $2N$ parties come together, they can
do a joint measurement to discriminate the states $\left\{ \rho_{2N-2}^{+}.\rho_{2N-2}^{-},\sigma_{2N-2}^{+},\sigma_{2N-2}^{-}\right\} $
(as they are mutually orthogonal, one would always be able to find
such measurements). Then it follows from Eqs. (\ref{rhoplusminus})
and (\ref{sigmaplusminus}), that they can create a maximally entangled
state between the remaining two parties via LOCC. This implies there
is one ebit of distillable entanglement across evcry $1:2N-1$ bipartite
partition. $$

In what follows, we show that \emph{$N$ ebits are both necessary
and sufficient} to prepare a $2N$ ($N\geq2$) qubit $\rho_{2N}^{+}$
state. 

\begin{center}\textbf{A Lower Bound on Entanglement Cost}\end{center}

In our model, every state preparation starts from a quantum resource
state of the form: $\rho_{B}=\otimes\prod_{i<j}^{N}\sigma_{ij}$,
where, the index $ij$ refers to the pair of parties $(i,j)$ sharing
the state $\sigma_{ij}.$ We begin with $2N$ spatially separated
nodes sharing such a resource state where $e_{ij}=e_{ji}$, $i,j\in\left\{ A_{1},A_{2},A_{3},...,A_{2N}\right\} $
and $i\not=j$, be the bipartite distillable entanglement (measured
in ebits) present between two parties $A_{i},A_{j}$. In the state
$\rho_{2N}^{+}$, consider the $2N,$ $1:2N-1$ bipartite cuts like,
$A_{k}:\{ A_{i,}i\neq k\}$. Across each one of these cuts, one can
distill one ebit of entanglement. Since LOCC can never increase distillable
entanglement, or amount of entanglement spent across a cut in preparing
the state should always be equal or more than the amount of distillable
entanglement across that cut, then we must have for a cut like $A_{k}:\{ A_{i,}i\neq k\}$,\begin{equation}
\sum_{i,i\not=k}e_{A_{k}A_{i}}\geq1\end{equation}
 We get one such inequality from each cut, corresponding to every
party, and if we sum them up, then we get \begin{equation}
\sum_{i,k,i\not=k}e_{A_{k}A_{i}}\geq2N\end{equation}
 Since, $e_{A_{k}A_{i}}=e_{A_{i}A_{k}},$we have,\begin{equation}
E=\sum_{k<i}e_{A_{k}A_{i}}\geq N,\label{eq:lb}\end{equation}
 where $E$ in the unit of ebits is the \textit{total bi-partite entanglement
shared between all the parties}. One might be tempted to argue that
the above bound has been derived from a single copy and hence, is
not an asymptotic bound. However, as stated in \emph{Remark 1}, any
asymptotic manipulation can only increase the distillable entanglement
of one ebit (as obtained from a single copy manipulation) across any
$1:(2N-1)$ cut, and hence the lower bound in Eq.~(\ref{eq:lb})
is a valid lower bound. 

\begin{center}\textbf{A local protocol achieving the lower bound }\end{center}

We now give a protocol that utilizes \textit{N} pairs of singlets
and LOCC to prepare $\rho_{2N}^{+}$. In fact, we show that a single
copy of the original state, where $N$ singlets are shared by $N$
disjoint pairs, is enough to prepare a single copy of the BCABE states,
and no asymptotic manipulation is necessary to achieve the lower-bound
derived above. The proof that our protocol indeed works can be seen
via induction. To begin with consider the state of four qubits, say
A, B, C and D. The following state was first presented by Smolin \cite{smolin00}
and corresponds to our class when $N=2$. 

\begin{eqnarray}
\rho_{ABCD}^{+} & = & \frac{1}{4}(\left[\Phi^{+}\right]_{AB}\otimes\left[\Phi^{+}\right]_{CD}+\left[\Phi^{-}\right]_{AB}\otimes\left[\Phi^{-}\right]_{CD}+\left[\Psi^{+}\right]_{AB}\otimes\left[\Psi^{+}\right]_{CD}\nonumber \\
 & + & \left[\Psi^{-}\right]_{AB}\otimes\left[\Psi^{-}\right]_{CD})\label{eq:}\end{eqnarray}

Let the pairs, (A, B), and (C, D), share a singlet each. A and C can
classically communicate among themselves to prepare a state $\left|\Phi_{i}\right\rangle ^{AA}\otimes\left|\Phi_{i}\right\rangle ^{CC}$
randomly with equal probability. This can be done as follows: Assume
that A and C, each of them possesses a Bell state generator. The generators
however generate identical Bell states randomly based on a string
of classical bits that can be established a priori. A and C then can
each teleport one qubit of the correlated Bell states (keeping one
qubit from each state to themselves) to B and D respectively using
the shared singlets. This creates the state $\rho_{ABCD}^{+}$. Suppose
now we have two additional parties E and F and all three pairs (A,
B), (C, D) and (E, F) share a singlet among thems and they would like
to prepare the following six qubit state: 

\begin{eqnarray}
\rho_{ABCDEF}^{+} & = & \frac{1}{4}(\left[\Phi^{+}\right]_{EF}\otimes\rho_{ABCD}^{+}+\left[\Phi^{-}\right]_{EF}\otimes\rho_{ABCD}^{-}+\left[\Psi^{+}\right]_{EF}\otimes\sigma_{ABCD}^{+}\nonumber \\
 & + & \left[\Psi^{-}\right]_{EF}\otimes\sigma_{ABCD}^{-})\label{eq:}\end{eqnarray}

Let us further note that the three other bound entangled states belonging
to the same class can be obtained by applying an appropriate local
pauli rotation on any one of the qubits. To begin with, A, B, C and
D will prepare a four qubit state as described before. Suppose E has
a Bell state generator and A has a machine that can apply a Pauli
rotation on the qubit. Furthermore, the machines share a common random
two bit classical string. Based on that classical string E's machine
generates a Bell state and accordingly A's machine applies a Pauli
rotation on the qubit of A. Once E teleports the qubit via the singlet
shared with F, the six party state is produced among them. 

It is obvious that the above strategy can be inductively extended
to any number of parties $2N$, for any $N$ greater than three. 

$$

\section{\textbf{Discussions}}

As argued before. this result shows an interesting feature, i.e.,
to prepare a mixed state in a multiparty setting even by an asymptotic
manipulation one may not do better than the single copy preparation.
This is in contrast with bipartite entanglement manipulation where
a mixed state preparation is necessarily more efficient asymptotically. 

$$ In Ref. \cite{Nielsen00} entanglement
of creation was introduced as the number of qubits per copy exchanged
between the parties to prepare the entangled state optimally. It was
also shown that entanglement of creation is equal to entanglement
of formation \cite{wootters98} for bipartite systems. For the bound
entangled states considered in the work, let us now point out the
equivalence of our concept of entanglement cost with that of entanglement
of creation. First note that the distillable entanglement across any
cut, as measured in ebits, is always a lower bound on the number of
qubis that need to be exchanged across the same cut. Hence, the lower
bounds derived here are also lower bounds for the entanglement of
creation. Next, the pairwise entanglements we used in our constructive
protocols are all singlets. A singlet can be always established by
sending a qubit. Hence, the states can be prepared by exchanging the
same number of qubits as the number of singlets used in our preparations.
Since the number of singlets match the lower bounds, the actual number
of qubits required to prepare the states also equal the lower bounds.
Thus, the exact entanglement of creation of our $2N$-party state
is also $N$. 

While preparing a single copy of the state, we showed how it can be
done by using singlets shared between $N$ pairs. This of course not
the only local way to prepare such state. Take for instance the four
qubit Smolin unlockable bound entangled state \cite{smolin00} which
belongs to our class of states when $N=2$. Instead of providing two
disjoint pairs with two singlets, one can think of a square configuration
where every edge has distillable entanglement equal to 0.5. In such
a distribution, the Smolin state, can be manufactured with an efficiency
of 2 ebits per copy only asymptotically. A similar efficiency can
also be achieved providing every pair with states having distillable
entanglement equal to 1/3. One should note that one can now in principle
assign states with varying distillable entanglement between the parties
but such a distribution would necessarily be inefficient in the sense
the cost of preparation per copy would go up. Let us emphasize that
only by providing singlets between the pairs one can achieve the optimal
value for a single copy preparation. 

Our approach has some obvious weaknesses. It relies heavily on the
knowledge of exact distillable entanglement across all bipartite partitions
of the multipartite state which are in general very difficult to compute.
Considering an extreme case where our approach fails is to compute
the entanglement cost of bound entangled states that are not activable.
Our partitioning argument does not work because across every partition
the state has zero distillable entanglement. However in many cases
it can be computed like our's for example and in such situations it
might be able to provide a good lower bound. $$

\emph{\large Acknowledgements}\\
 One of the authors (S.B) is thankful to W. K. Wootters for many helpful
comments on this work. The work of S.B. was supported by iCORE, MITACS,
General Dynamics Canada, and CIAR. This work was sponsored in part
by the Defense Advanced Research Projects Agency (DARPA) project MDA972-99-1-0017,
and in part by the U. S. Army Research Office/DARPA under contract/grant
number DAAD 19-00-1-0172.

\end{document}